# A high-κ wide-gap layered dielectric for two-dimensional van der Waals heterostructures

Aljoscha Söll[1*], Edoardo Lopriore[2,3*], Asmund K. Ottesen[2,3], Jan Luxa[1], Gabriele Pasquale[2,3], Jiri Sturala[1], František Hájek[4], Vítězslav Jarý[4], David Sedmidubský[1], Kseniia Mosina[1], Andras Kis[2,3], Zdeněk Sofer[1§]

[1]*Department of Inorganic Chemistry, University of Chemistry and Technology Prague, Technicka 5, 166 28 Prague 6, Czech Republic*
[2]*Institute of Electrical and Microengineering, École Polytechnique Fédérale de Lausanne (EPFL), CH-1015 Lausanne, Switzerland*
[3]*Institute of Materials Science and Engineering, École Polytechnique Fédérale de Lausanne (EPFL), CH-1015 Lausanne, Switzerland*
[4] *Institute of Physics of the Czech Academy of Sciences, v.v.i., Cukrovarnická 10, 162 00, Prague 6, Czech Republic*

*\* These authors contributed equally to this work.*
*§ Correspondence should be addressed to: Zdenek Sofer (zdenek.sofer@vscht.cz)*

**ABSTRACT**

**Van der Waals heterostructures of two-dimensional materials have opened up new frontiers in condensed matter physics, unlocking unexplored possibilities in electronic and photonic device applications. However, the investigation of wide-gap high-κ layered dielectrics for devices based on van der Waals structures has been relatively limited. In this work, we demonstrate an easily reproducible synthesis method for the rare earth oxyhalide LaOBr, and we exfoliate it as a 2D layered material with a measured static dielectric constant of $\epsilon_{0,\perp} \simeq 9$ and a wide bandgap of 5.3 eV. Furthermore, our research demonstrates that LaOBr can be used as a high-κ dielectric in van der Waals field-effect transistors with high performance and low interface defect concentrations. Additionally, it proves to be an attractive choice for electrical gating in excitonic devices based on 2D materials. Our work demonstrates the versatile realization and functionality of 2D systems with wide-gap and high-κ van der Waals dielectric environments.**

**INTRODUCTION**

The International Roadmap for Devices and Systems (IRDS) has identified two-dimensional semiconductors as the channel materials for future technology nodes in the coming decade[1]. 2D materials have been proven to be compatible with standard semiconductor fabrication methods including the CMOS process[2]. Moreover, they have opened up numerous possibilities in non-computational systems under the more-than-Moore paradigm, ranging from back-end-of-the-line (BEOL) to co-integration with silicon technology in monolithic solutions[3].

The scaling of silicon technology has led to the transition of the insulating materials for electrical gating in transistor devices to high-κ dielectrics. However, the integration of 2D materials with 3D high-κ dielectrics, such as $Al_2O_3$ and $HfO_2$, is challenging due to the presence of dangling bonds at their interface, which adversely impact device performance[4]. Even though high temperature treatment or the use seeding layers have resulted in improved performance of electrical devices based on atomic-layer-deposited (ALD) oxides[5], the achievement of ideal interfaces remains a work in progress.

Native high-κ oxides have been proven to offer ideal dielectric interfaces[6], but their application is limited to semiconductors that allow for native oxide growth. High-κ perovskite films have also been transferred onto 2D materials, demonstrating gate dielectric functionality with van der Waals interfaces[7,8]. However, these alternatives lack the modularity enabled by fully van der Waals heterostructures. In fact, a key strenght of using 2D materials for device fabrication is the possibility of stacking them into heterostructures without critical structural constraints[9]. This has paved the way for exploring unprecedented electrical, optical, and emergent physical effects thanks to the combination of different 2D materials in freely configurable stacks. Hexagonal boron nitride (hBN) has played a crucial role in these advancements as the primary layered dielectric to be successfully employed in van der Waals heterostructures of 2D materials. However, the static dielectric constant of hBN is limited to values lower than that of $SiO_2$[4]. Therefore, high-κ layered dielectrics are required as an essential step toward the future scaling of a fully-2D MOSFET platform.

Furthermore, the use of high-κ dielectrics as encapsulation layers in van der Waals heterostructures would enable unexplored optoelectronic device applications. As an example, in the case of excitonic properties of transition metal dichalcogenides (TMDCs), dipolar repulsive interactions in the transport of spatially indirect inter-layer excitons are expected to be enhanced in high-κ dielectric environments, providing scaling possibilities for the long-range propagation of exciton ensembles[10].

Recently, $Bi_2SeO_5$ has been synthesized as a single crystal high-κ layered material for use as an insulator[11]. However, its low bandgap of $E_g \simeq 3.6$ eV critically limits its possible applications as a versatile dielectric in 2D field-effect structures, since its $E_g$ is comparable to that of materials such as InSe, GaSe and GaS[12]. In fact, in the search for high-κ dielectrics, versatile gate insulators require a band alignment with barriers of 1eV for both electrons and holes[13]. Considering that most TMDC bandgaps are around 2 eV, and that metal monochalcogenides exhibit gaps up to 3 eV[12], a van der Waals high-κ dielectric with $E_g > 5$ eV is highly desirable to enable wide material-independent van der Waals integration in 2D heterostructures. Rare earth oxyhalides have been theoretically predicted to be easily exfoliable layered insulators with wide bandgap and high-κ both in bulk and monolayer forms[14]. In this work, we synthesize crystals of LaOBr, a rare earth oxybromide that has been indicated as an ideal layered dielectric for 2D heterostructure devices offering low leakage currents and a wide bandgap[14]. Here, we develop a high-temperature flux growth method for LaOBr that focuses on the production of crystals in the form of platelets, and we provide its complete bulk characterization, highlighting its bulk bandgap of approximately 5.3 eV.

We reveal the layered structure of LaOBr, exfoliate it, and employ the established pick-and-place fabrication techniques to build van der Waals heterostructures with LaOBr as a dielectric material. By conducting electrical transport measurements in graphene field-effect structures, we determine the static out-of-plane dielectric constant to be $\epsilon_{0,\perp} \simeq 9$. Subsequently, we employ LaOBr as a gate dielectric for transistor operation in a field-effect structure with MoS$_2$. We observe a low subthreshold slope ($\sim 85\,\mathrm{mV\,dec^{-1}}$) and a corresponding low interface defect concentration ($D_{it} \simeq 1.06 \cdot 10^{12}\,\mathrm{cm^{-2}eV^{-1}}$), together with a high on-off ratio ($I_{\mathrm{on}}/I_{\mathrm{off}} > 10^8$), low leakage current ($< 10^{-4}\,\mathrm{A\,cm^{-2}}$), and negligible I-V hysteresis. These results provide the evidence that LaOBr can be used as a high-κ dielectric in high-performance fully van der Waals electronic devices.

Furthermore, we demonstrate that LaOBr can be employed as encapsulation and gate insulator for excitonic devices based on 2D materials, showing the modulation of excitonic species in MoSe$_2$ by electronic gating. This study highlights the promising potential of LaOBr in advancing the field of 2D heterostructures and their applications in electronic devices.

## RESULTS

### LaOBr crystal synthesis

Single-crystal growth of rare earth oxyhalides has been scarcely documented in the literature until now. In particular, the synthesis of LaOBr in crystal form has been previously achieved using anhydrous LaBr$_3$ and La$_2$O$_3$ with an excess of LaBr$_3$ as a flux, leading primarily to the formation of LaOBr as a mixture of needles and narrow platelets[15]. Although we were able to successfully replicate this growth (Supplementary Note 1), the predominance of needles is problematic for van der Waals heterostructure fabrication. Notably, needles of LaOBr tend to grow perpendicularly to the c-plane, likely along a screw dislocation, resulting in a minimal (001) cross-section, making them unsuitable for exfoliation (Supplementary Fig. 1). Furthermore, the previously developed synthesis is performed in a sealed quartz ampule, which significantly reduces its accessibility and introduces potential hazards stemming from gas production during the synthesis.

In order to devise a more suitable synthesis of LaOBr (Fig. 1a-b), we conducted thermogravimetric analysis (TGA) coupled with differential scanning calorimetry (DSC) on $LaBr_3 \cdot 7\ H_2O$ in dynamic argon atmosphere. The resulting TGA is shown in Fig. 1c, and a more detailed description is given in Supplementary Note 2. The TGA-DSC analysis indicates that, contrary to the procedure established by Haeuseler et al.[13], we do not require the use of anhydrous $LaBr_3$ in our method, as the crystal water is eliminated during the reaction when we use an open vessel. Moreover, this procedure eliminates the need for adding $La_2O_3$ since atmospheric oxygen is sufficient as the main source of oxygen (Supplementary Fig. 2).

Therefore, our modified LaOBr growth method takes place in an open corundum crucible under atmospheric conditions using an eutectic mixture of alkaline and earth-alkaline metal-salts. We chose a mixture of NaBr and $MgBr_2$ as a flux with a low melting point to maximize the temperature range for material growth. By avoiding quartz ampules, we further aim to increase the accessibility of LaOBr thanks to its improved growth reproducibility (Methods, Supplementary Note 3).

After performing the synthesis at 1000 °C and separating the material from the flux, we obtained LaOBr in the form of colorless platelets with lateral dimensions of up to 1 mm (Fig. 2a). Utilizing this method, we successfully eliminated the formation of needles, yielding only thin platelets of LaOBr that are highly suitable for exfoliation (Supplementary Fig. 3).

**LaOBr bulk characterization**

We verified the phase purity of the obtained LaOBr platelets using X-Ray powder diffraction (XRD), as shown in Fig. 1f. LaOBr crystallizes in the tetragonal space group P4/nmm with no additional phases. Its lattice constants were calculated from Rietveld refinement as $a = b = 4.1618\ \text{Å},$ and $c = 7.3813\ \text{Å}\ (\alpha = \beta = \gamma = 90°)$. Figure 1d shows the crystal structure obtained from the refinement, revealing a clear van der Waals gap perpendicular to the c-axis.

We note that LaOBr exhibits a tetragonal symmetry, an uncommon property among van der Waals materials, most of which possess a hexagonal symmetry[9,16]. Furthermore, a monolayer of the material comprises 5 rows of atoms, in contrast to TMDCs with 3 rows of atoms, or hBN and graphene which are a single atom thick.

As shown in Fig. 1e, we performed Raman spectroscopy on the bulk sample, assigning the Raman modes according to the literature, and confirming this assignment using Density Functional Theory (DFT) calculations (Supplementary Note 4). Furthermore, we have analyzed the surface composition by X-ray photoelectron spectroscopy (XPS), demonstrating the absence of surface impurities. The corresponding XPS spectra and obtained binding energies are discussed in detail in Supplementary Note 5.

We characterized the sample morphology using scanning electron microscopy (SEM), equipped with an energy dispersive X-ray spectrometer (EDS). Based on SEM measurements, we observed that LaOBr forms thin platelets with lateral dimensions ranging from 100 μm to 1 mm (Fig. 2a) and approximate thicknesses of around $5 - 20$ μm. Many of the platelets display a terraced morphology on their edges, which is typical for layered structures (Fig. 2b, Supplementary Fig. 3). The EDS analysis confirms the correct stoichiometry of the samples, with an atomic ratio of $33 \pm 1$ % for each element. Additionally, the EDS map taken from a single crystal of LaOBr reveals a uniform distribution of all three elements across the entire crystal (Fig. 2c).

To determine the bandgap of bulk LaOBr, we performed photoluminescence excitation spectroscopy (PLE), which yielded a value of 5.3 eV (Fig 2f). Further details can be found in Supplementary Note 6. This value is in agreemend with previous measurements using UV-Vis spectroscopy[17].

**Dielectric constant estimation**

Like other rare earth oxyhalides, LaOBr exhibits a significant difference between its optical and static dielectric constants[14]. Here, we focus on the optical and static dielectric constants in the out-of-plane direction, which are particularly relevant for identifying and using LaOBr as a layered gate dielectric. The out-of-plane optical dielectric constant, which considers only the electronic response, was calculated to be $\epsilon_{0,\infty} \sim 4.6$ for LaOBr[14]. This value is important for identifying LaOBr flakes on a substrate, since the optical refractive index of the material is given by its optical dielectric constant. Due to its bandgap of $E_g \sim 5.3\ eV$ and low optical dielectric constant of 4.6, LaOBr can be easily distinguished on SiO$_2$ and PDMS substrates, commonly used for the exfoliation of 2D materials. Moreover, since the optical dielectric constant of LaOBr is close to that of hBN[14], its thickness can be estimated by color contrast in a similar manner as it is done for hBN[18]. Figure 2d shows the identification of various thicknesses of LaOBr on SiO$_2$, with corresponding AFM measurements shown in Fig. 2e. LaOBr flakes tend to exhibit rectangular shapes due to their tetragonal crystal symmetry, as shown in Fig. 2d and in Supplementary Fig. 5. We note that the ease of exfoliation and identification of the 2D dielectric material, along with the ability to estimate its thickness by optical means, are crucial factors for successfully incorporating the high-κ dielectric in the fabrication of van der Waals heterostructures based on the most established substrates and techniques.

On the other hand, the static dielectric constant $\epsilon_{0,\perp}$, comprising both ionic and electronic responses, is the key reference value for electrical gating in field-effect structures. Specifically, the out-of-plane static dielectric constant of LaOBr has been theoretically predicted to be 12.5 in its bulk form, with a slightly higher value for the monolayer case[14]. To determine $\epsilon_{0,\perp}$ by electrical measurements, we utilize graphene field-effect dual-gated devices (Fig. 3a-b). We fabricate fully van der Waals field-effect heterostructures using an established dry pick-up technique, showcasing the exfoliatable and stackable nature of LaOBr as a 2D dielectric (Methods, Supplementary Fig. 6).

The Dirac point of graphene in field-effect structures corresponds to the position of maximum lateral resistance in gate sweeps with a fixed bias voltage (Fig. 3d). By performing dual-gate voltage sweeps, we can compare the capacitive field-effect modulation from the bottom and top gate dielectrics, represented by SiO$_2$ and LaOBr, respectively[8,19]. We use SiO$_2$ as an established reference insulator with $\epsilon_{0,\perp}^{SiO2} \sim 3.9$. Therefore, the slope of the change in the Dirac point in Fig. 3e is directly determined by the ratio of bottom and top gate capacitance:

$$\frac{C_{SiO_2}}{C_{LaOBr}} = \frac{\epsilon_{SiO2}\, t_{LaOBr}}{\epsilon_{LaOBr}\, t_{SiO2}} \tag{1}$$

Our bottom substrate oxide has a fixed thickness of $t_{SiO2} = 270$ nm, while the LaOBr flakes have a measured thickness of $t_{LaOBr,A} = 46$ nm and $t_{LaOBr,B} = 56$ nm, as dermined by AFM. The different slopes observed in the linear fit of the Dirac point shift are due to the varying thickness of LaOBr. From Eq. 1, we calculate the average out-of-plane static dielectric constant of LaOBr to be $\epsilon_{0,\perp} \simeq 9 \pm 0.4$ (Supplementary Fig. 7). This value aligns within 25% of the theoretically calculated value of 12.5, and the uncertainty range falls within the common margin of calculations for other low-κ van der Waals dielectrics[14]. With a measured bandgap of 5.3 eV and a static dielectric constant of 9, LaOBr is a 2D equivalent of commonly employed 3D oxides such as Al$_2$O$_3$ and Si$_3$N$_4$ (Supplementary Fig. 8), with the crucial feature of full integration with van der Waals systems.

Additionally, we measure the gate leakage current in our graphene field-effect devices with LaOBr, finding values that are orders of magnitude lower than both the gate limit and the low-power limit[13] for a range of electric fields up to 1.5 MV cm$^{-1}$ (Fig. 3c). Furthermore, we assessed the breakdown of our LaOBr flakes using metal-insulator-metal (MIM) structures, obtaining a consistent breakdown field value of $E_{BD} \simeq 8$ MV cm$^{-1}$ for different bulk thicknesses (Supplementary Fig. 9). The measured breakdown is comparable to that of hBN[20] and of other commonly used 3D bulk oxides as $Al_2O_3$[21]. These results demonstrate that LaOBr can be used as a high-κ and low-leakage layered dielectric material, which aligns well with theoretical predictions[14].

**Fully van der Waals high-κ field-effect transistor**

To showcase the applicaiton of LaOBr as a high-κ insulator for fully van der Waals devices, we fabricated a field-effect transistor using few-layer $MoS_2$ as the channel material and a 20 nm thick LaOBr as the gate dielectric (Supplementary Fig. 10). Figure 4a shows the room-temperature transfer characteristics obtained by sweeping the top-gate voltage $V_{TG}$ with a fixed bias $V_{DS}$. All sweeps are performed in both directions of the $V_{TG}$ scale. In the inset of Figure 4a, we focus on the forward sweep with $V_{DS} = 200\ mV$, revealing a threshold voltage of $V_{TH} \simeq -2\ V$ and a field-effect two-terminal mobility of $\mu_{FE} \simeq 32\ cm^2\ V^{-1}\ s^{-1}$. The obtained mobility is comparable with the previously reported two-terminal $\mu_{FE}$ obtained for few-layer $MoS_2$ with lateral few-layer graphite contacts[8].

Hysteretic behaviors in transfer curves are commonly observed in 2D material field-effect transistors based on bulk 3D oxides[22]. These non-idealities result from high defect densities introduced by dangling bonds at oxide-semiconductor interfaces. However, in our case, all transfer curves in Fig. 4a exhibit negligible hysteresis in the order of tens of mV, indicating a low defect density at the interface between LaOBr and $MoS_2$. Moreover, we measured the output characteristics of the LaOBr/$MoS_2$ transistor by sweeping the bias voltage $V_{DS}$ with a fixed $V_{TG}$ (Fig. 4b-c). Figure 4b shows a highly linear behavior at low bias voltages independent of the gate modulation, indicating the absence of Schottky barrier effects. The field-effect modulation of the channel conductance reveals the conventional linear and saturation regimes of 2D MOSFET devices (Fig. 4c). Additionally, in Fig. 4d, we show the transfer curves in logarithmic scale to highlight the subthreshold trend, and we observe a subthreshold slope as low as 85 mV/dec. This value is directly related to the interface defect density $D_{it}$ by $SS = \frac{kT}{q}\log\left(1 + \frac{qD_{it}}{C_{ox}}\right)$, and we estimate it to be $D_{it} \simeq 1.06 \cdot 10^{12}$ cm$^{-2}$eV$^{-1}$, which is comparable to other high-performing devices based on MoS$_2$[4]. Therefore, the negligible hysteresis and the low subthreshold slope both indicate an ideal van der Waals interface between LaOBr and MoS$_2$.

**Exciton control with LaOBr**

The electrical control of excitonic features in optically-active 2D materials has significant implications for both fundamental discoveries and novel applications in the field of optoelectronics[23,24]. In particular, van der Waals dielectric encapsulation provides protection and enables electrical gating of active TMDCs, facilitating the observation of charged excitonic species known as trions. Here, we fabricated a heterostructure with monolayer MoSe$_2$ fully encapsulated by LaOBr on a bottom metal gate (Fig. 5a-b). In Fig. 5c-d, we present the gate-dependent excitonic features in our device, measured by photoluminescence (PL) spectroscopy at a temperature of 4 K (Methods). The presence of an impurity-bound state $X_I$ is aligned with what is generally obtained for high-quality TMDCs on different substrates[25,26].

We note that the linewidth of the obtained excitonic species is in the order of tens of meV, comparable to that obtained on SiO$_2$ at cryogenic temperatures. It is known that inhomogeneous broadening of excitons in TMDCs is mainly caused by local variabilities in the dielectric environment. Thus, the presence of inhomogeneous broadening in our features suggests that the dielectric is not perfectly conformal to the TMDC. We attribute this to the aforementioned complex molecular structure of LaOBr (Fig. 1d). Nevertheless, we demonstrated that LaOBr can be used as an encapsulation dielectric and as a gate dielectric to control excitonic features in 2D materials, opening up a new high-κ playground for excitonic devices based on van der Waals heterostructures.

**DISCUSSION**

We have successfully developed a straightforward and reproducible high-temperature flux growth method for producing large, stoichiometric crystals of the high-κ dielectric LaOBr. Through comprehensive characterization, we examined LaOBr both in its bulk form and as an exfoliated flake, utilizing it as the insulating material in van der Waals heterostructure devices. With an out-of-plane static dielectric constant of 9, a high bandgap (5.3 eV), robust dielectric breakdown (8 MV cm$^{-1}$) and low leakage currents ($< 10^{-4}$ A cm$^{-2}$), LaOBr is a near-perfect dielectric for gating and encapsulation of 2D materials.

Thanks to the exceptional quality of our LaOBr crystals and their ease of integration into van der Waals assembly processes, we successfully demonstrated field-effect transistor action on few-layer MoS$_2$ with negligible hysteresis, low subthreshold slope ($\sim 85$ mV dec$^{-1}$) and a low interfacial defect concentration ($D_{it} \simeq 1.06 \cdot 10^{12}$ cm$^{-2}$eV$^{-1}$). These results validate LaOBr as a layered, van der Waals-compatible, high-κ dielectric with a high bandgap and excellent breakdown voltage, representing a pivotal advancement towards the future scaling of electronics based on 2D materials.

Furthermore, our investigation showcased the utility of LaOBr as an encapsulating dielectric for studying gate-dependent excitonic features in a van der Waals heterostructures, opening venues towards excitonic devices with high-κ dielectric environments.

In the context of dielectric scaling over the past decades, LaOBr emerges as a van der Waals alternative to bulk oxides like Al$_2$O$_3$, offering the critical advantage of its layered structure. Utilizing LaOBr as a high-κ wide-gap layered dielectric for 2D material-based devices lays the foundation for exploring electrical and optical devices within a high-κ dielectric environment, without any limitations on material combinations, thus exemplifying the paradigm of fully van der Waals integration.

**ACKNOWLEDGEMENTS**

We acknowledge the support in microfabrication and e-beam lithography from EPFL Centre of MicroNanotechnology (CMI) and thank Z. Benes (CMI) for help with electron-beam lithography. This work was financially supported by the Marie Curie Sklodowska ITN network "2-Exciting" (grant no. 956813). We also acknowledge the computational resources, which were provided by the e-INFRA CZ project (ID:90254), supported by the Ministry of Education, Youth and Sports of the Czech Republic. This work was also supported by ERC-CZ program (project LL2101) from the Ministry of Education, Youth and Sports of the Czech Republic (MEYS).

## METHODS

**Crystal synthesis**

As a starting material, lanthanum carbonate octahydrate (99.95%, Ganzhou Wanfeng Adv. Materials Tech. Co., Ltd., China) was used without any further purification. $LaBr_3 \cdot 7 H_2O$ was obtained by dissolving the oxide in concentrated hydrobromic acid (47%, p.a. grade, Fisher Scientific, Czech Republic) and crystallization of heptahydrate from acidic solution (pH ~5) on a steam bath. Mixtures of 5 g $LaBr_3 \cdot 7 H_2O$ (made from lanthanum carbonate and hydrobromic acid), 4.2 g NaBr (99%, p.a. grade, LachNer, Czech Republic), and 10.9 g $MgBr_2$ (98%, Sigma Aldrich, Czech Republic) were placed in a corundum crucible. The crucible was initially heated to 150°C for 5 hours, where the majority of water slowly evaporates according to TGA. We placed the furnace in a well-ventilated area (fume hood) because of HBr gas production during this step. After dehydration, the mixture was slowly heated to 1000°C at 2°C min-1, the temperature was maintained for 48 hours, and subsequently slow cooled down to 650°C at 0.1°C min-1, during which the growth of LaOBr takes place. After reaching 650°C, the reaction was free-cooled to room temperature. The corundum crucible was leached in boiling water for 24h to dissolve the salt flux, and the insoluble product was separated through vacuum filtration. The use of MgBr$_2$ as a flux introduced a small amount of water-insoluble MgO which was removed by washing with dilute sulfuric acid (1:10, analytical reagent grade). The product was obtained as colorless platelets of LaOBr with lateral dimensions up to 1mm.

**Bulk LaOBr characterization**

X-ray diffraction was carried out on a Bruker D8 Dis-cover with Cu X-ray source ($\lambda$= 0.15418 nm, U= 40 kV, I= 40 mA). Diffractograms were collected in a range from 10° to 90° range with a step of 0.02° and integration time of 0.2s. The data was processed in HighScore plus software package and the Rietveld Refinement was conducted in Fullprof. Diffractograms were then normalized to the most intense peak. Characterization by Atomic Force Microscopy (AFM) was performed on NT-MDT Ntegra Spectra from NT-MDT in tapping mode and on the Asylum Research Cypher system. The morphology of samples was investigated using scanning electron microscopy (SEM) with a FEG electron source (Tescan Lyra dual beam microscope). The samples were placed on a carbon conductive tape. SEM measurements were carried out using a 5 kV electron beam. The composition of the samples was determined using an energy dispersive spectroscopy (EDS) analyzer (X-MaxN) with a 20 mm$^2$ SDD detector (Oxford Instruments). Data was evaluated using AZtecEnergy software. EDS measurements were carried out with a 15 kV acceleration voltage. Raman spectra were recorded with a WITec Confocal Raman Microscope (WITec alpha300 R, Ulm, Germany), equipped with a 532 nm laser and a spectrometer with a thermoelectrically cooled CCD camera sensor. The measurement was performed at room temperature with a 100× objective and a laser power of less than 1.2 mW to avoid sample degradation. The Raman modes were assigned according to DT calculations in Quantum Espresso (Supplementary Note 4). Photoluminescence excitation (PLE) spectra were measured by a custom-made spectrofluorometer 5000M (Horiba Jobin

Yvon, Wildwood, MA, USA) using steady state laser driven xenon lamp (Energetiq, a Hamamatsu Company) as the excitation sources. The detection part of the setup involved a single-grating monochromator and a photon-counting detector TBX-04 (Hamamatsu). The PLE spectra are corrected for the experimental distortion. Thermogravimetric experiments (TG) were performed on a Themys TGA (SETARAM instrument) with a heating rate of 10 °C min$^{-1}$. The instrument was purged for at least three hours by the carrier gas.

**Device fabrication**

All devices used in this work were fabricated on top of SiO$_2$(270 nm)/p+ Si substrates. For graphene field-effect structures, graphite (NGS) was exfoliated directly on the substrate and graphene was identified by AFM. LaOBr was exfoliated on PDMS (gelpak) and transferred on graphene by a dry technique. In particular, LaOBr was picked up from PDMS with a polycarbonate (PC) membrane on a PDMS stamp. Then, LaOBr was transferred with the PC membrane on top of graphene. For MoS$_2$ field-effect structures, neighboring exfoliated few-layer graphite flakes were used as lateral electrodes. MoS$_2$ (2D Semiconductors) flakes were exfoliated on PDMS (gelpak). LaOBr was picked up from PDMS and then used to pick up MoS$_2$. The MoS$_2$/LaOBr stack was then transferred on top of the few-layer graphite flakes used as lateral drains and sources. All samples were annealed in high vacuum (10$^{-6}$ mbar) at 340 °C for 6 hours. Top gates and contacts were all fabricated by e-beam lithography and metal evaporation (2 nm/80 nm Ti/Au).

**Electrical measurements**

Transport measurements were carried out at room temperature under high vacuum (10$^{-6}$ mbar) with dual-channel Keithley 2636 source measure units (SMUs). In all devices, a lateral contact was kept grounded, while the other lateral contact and the top gate were connected to the two channels of the same SMU. For dual-gate graphene field-effect structures, the silicon backgate was connected to another Keithley 2636 SMU.

**Optical measurements**

Optical measurements were performed in a vacuum in a He-flow cryostat at 4.6 K. Excitons in $MoSe_2$ were excited with a confocal microscope by a continuous-wave 647 nm diode laser focused to the diffraction limit, and the emitted photons were collected through the same objective. The laser spot full-width at half-maximum measured approximately 1.2 µm. The emitted light was filtered by a 650 nm long-pass edge filter and then acquired using a spectrometer (Andor Shamrock) and recorded with a CCD (charge-coupled device) camera (Andor Newton).

**AUTHOR CONTRIBUTIONS**

Z.S. and A.K. initiated and supervised the project. A.S. and Z.S. grew the LaOBr crystals. E.L., A.S. and A.O. fabricated the devices, assisted by G.P.. E.L. performed the $MoS_2$ field-effect transistor measurements, assisted by G.P.. E.L. analyzed the data with input from A.S. and A.K.. J.L performed XPS measurements and analyzed the data. J.S. performed TGA measurements and DFT calculations. F.H. and V.J performed PLE measurements and F.H. analyzed the data. D.S. performed Rietveld Refinement. E.L. and K.M. performed AFM measurements. E.L. and A.S. wrote the manuscript, with inputs from all authors.

**COMPETING FINANCIAL INTERESTS**

The authors declare no competing financial interests.

## DATA AVAILABILITY

The data that support the findings of this study are available from the corresponding author on reasonable request.

## REFERENCES

1. International Roadmap for Devices and Systems (IRDS$^{TM}$) 2021 Edition - IEEE IRDS$^{TM}$. https://irds.ieee.org/editions/2021.

2. Akinwande, D. *et al.* Graphene and two-dimensional materials for silicon technology. *Nature* **573**, 507–518 (2019).

3. Lemme, M. C., Akinwande, D., Huyghebaert, C. & Stampfer, C. 2D materials for future heterogeneous electronics. *Nat Commun* **13**, 1392 (2022).

4. Illarionov, Y. Y. *et al.* Insulators for 2D nanoelectronics: the gap to bridge. *Nat Commun* **11**, 3385 (2020).

5. Zou, X. *et al.* Interface Engineering for High-Performance Top-Gated MoS2 Field-Effect Transistors. *Advanced Materials* **26**, 6255–6261 (2014).

6. Li, T. *et al.* A native oxide high-κ gate dielectric for two-dimensional electronics. *Nat Electron* **3**, 473–478 (2020).

7. Huang, J.-K. *et al.* High-κ perovskite membranes as insulators for two-dimensional transistors. *Nature* **605**, 262–267 (2022).

8. Yang, A. J. *et al.* Van der Waals integration of high-κ perovskite oxides and two-dimensional semiconductors. *Nat Electron* **5**, 233–240 (2022).

9. Novoselov, K. S., Mishchenko, A., Carvalho, A. & Neto, A. H. C. 2D materials and van der Waals heterostructures. *Science* **353**, aac9439 (2016).

10. Erkensten, D., Brem, S., Perea-Causín, R. & Malic, E. Microscopic origin of anomalous interlayer exciton transport in van der Waals heterostructures. *Phys. Rev. Materials* **6**, 094006 (2022).

11. Zhang, C. *et al.* Single-crystalline van der Waals layered dielectric with high dielectric constant. *Nat. Mater.* 1–6 (2023) doi:10.1038/s41563-023-01502-7.

12. Chaves, A. *et al.* Bandgap engineering of two-dimensional semiconductor materials. *npj 2D Mater Appl* **4**, 1–21 (2020).

13. Robertson, J. High dielectric constant gate oxides for metal oxide Si transistors. *Rep. Prog. Phys.* **69**, 327 (2005).

14. Osanloo, M. R., Van de Put, M. L., Saadat, A. & Vandenberghe, W. G. Identification of two-dimensional layered dielectrics from first principles. *Nat Commun* **12**, 5051 (2021).

15. Haeuseler, H. & Jung, M. Single crystal growth and structure of LaOBr and SmOBr. *Materials Research Bulletin* **21**, 1291–1294 (1986).

16.     Manzeli, S., Ovchinnikov, D., Pasquier, D., Yazyev, O. V. & Kis, A. 2D transition metal dichalcogenides. *Nature Reviews Materials* **2**, 1733 (2017).

17.     Kim, D., Park, S., Kim, S., Kang, S.-G. & Park, J.-C. Blue-Emitting Eu2+-Activated LaOX (X = Cl, Br, and I) Materials: Crystal Field Effect. *Inorg. Chem.* **53**, 11966–11973 (2014).

18.     Anzai, Y. *et al.* Broad range thickness identification of hexagonal boron nitride by colors. *Appl. Phys. Express* **12**, 055007 (2019).

19.     Xu, H. *et al.* Quantum Capacitance Limited Vertical Scaling of Graphene Field-Effect Transistor. *ACS Nano* **5**, 2340–2347 (2011).

20.     Hattori, Y., Taniguchi, T., Watanabe, K. & Nagashio, K. Layer-by-Layer Dielectric Breakdown of Hexagonal Boron Nitride. *ACS Nano* **9**, 916–921 (2015).

21.     Cheng, L. *et al.* Sub-10 nm Tunable Hybrid Dielectric Engineering on MoS2 for Two-Dimensional Material-Based Devices. *ACS Nano* **11**, 10243–10252 (2017).

22.     Bartolomeo, A. D. *et al.* Hysteresis in the transfer characteristics of MoS2 transistors. *2D Mater.* **5**, 015014 (2017).

23.     Wang, Q. H., Kalantar-Zadeh, K., Kis, A., Coleman, J. N. & Strano, M. S. Electronics and optoelectronics of two-dimensional transition metal dichalcogenides. *Nat Nano* **7**, 699–712 (2012).

24.     Ciarrocchi, A., Tagarelli, F., Avsar, A. & Kis, A. Excitonic devices with van der Waals heterostructures: valleytronics meets twistronics. *Nat Rev Mater* (2022) doi:10.1038/s41578-021-00408-7.

25.     Ross, J. S. *et al.* Electrical control of neutral and charged excitons in a monolayer semiconductor. *Nat Commun* **4**, 1474 (2013).

26.     Rivera, P. *et al.* Intrinsic donor-bound excitons in ultraclean monolayer semiconductors. *Nat Commun* **12**, 871 (2021).

# FIGURES

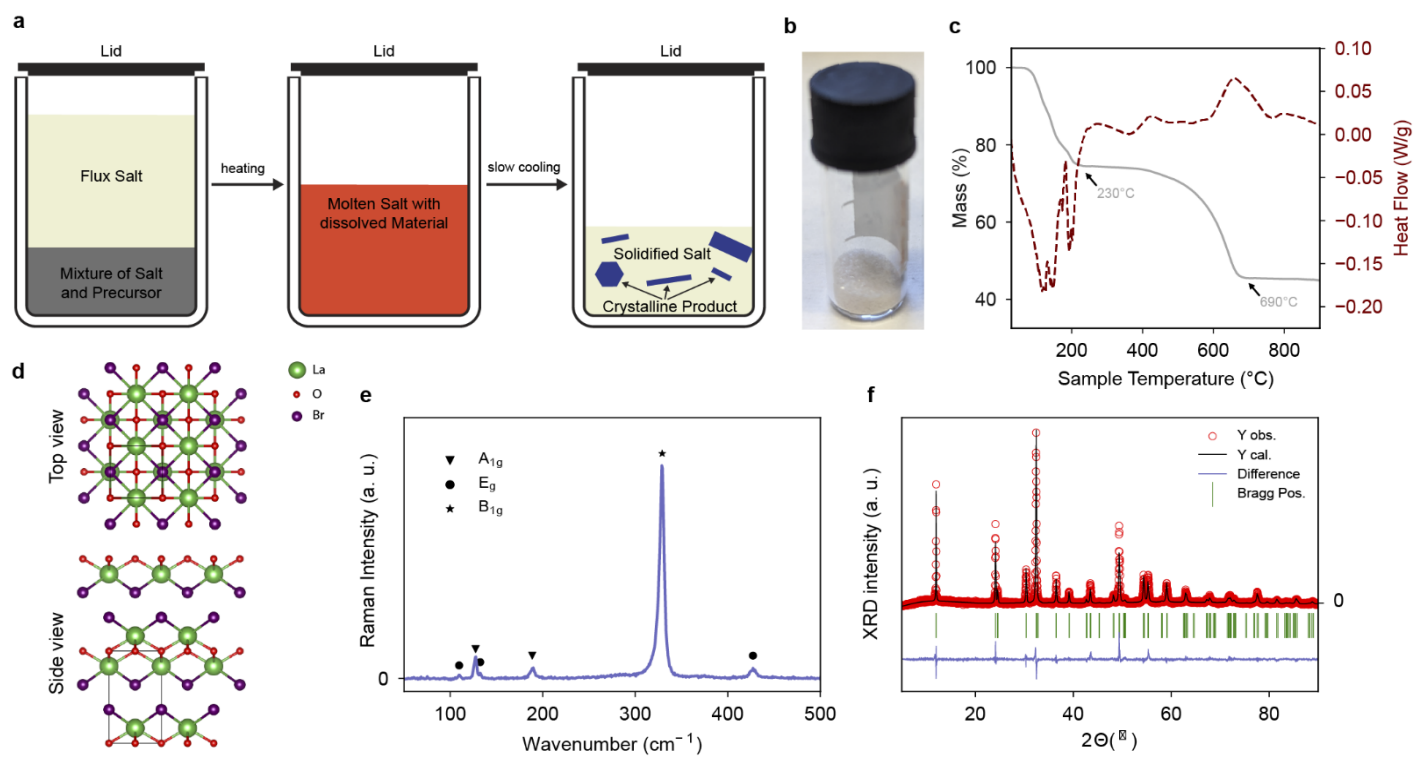

**Figure 1. Synthesis and characterization of bulk LaOBr. a,** Synthesis schematic for the growth of LaOBr single crystals for the method used in this work. **b,** LaOBr bulk crystals obtained from the synthesis. **c,** TGA-DSC measurements of $LaBr_3 \cdot 7\,H_2O$ in $Ar/O_2$ atmosphere shows gradual conversion into anhydrous $LaBr_3$ completed at 230°C, and finally LaOBr completed at 690°C. **d,** Refined structure of LaOBr as obtained from XRD, visualized with VESTA[27]. **e,** Raman spectroscopy on bulk LaOBr. We assign the two $A_{1g}$ modes to 127.3 cm$^{-1}$ ($A_{1g1}$) and 188.5 cm$^{-1}$ ($A_{1g2}$). We assign the three $E_g$ modes to 109.5 cm$^{-1}$ ($E_{g1}$), 132.3 cm$^{-1}$ ($E_{g2}$), and 427.1 cm$^{-1}$ ($E_{g3}$), and the intense ($B_{1g}$) mode to 329.0 cm$^{-1}$ ($B_{1g1}$), based on DFT calculations (Supplementary Note 4) and in accordance with the literature[28,29]. **f,** X-ray powder diffractogram (XRD) of bulk LaOBr with experimental data in red, Rietveld refinement fit in black and literature peak positions in green[15].

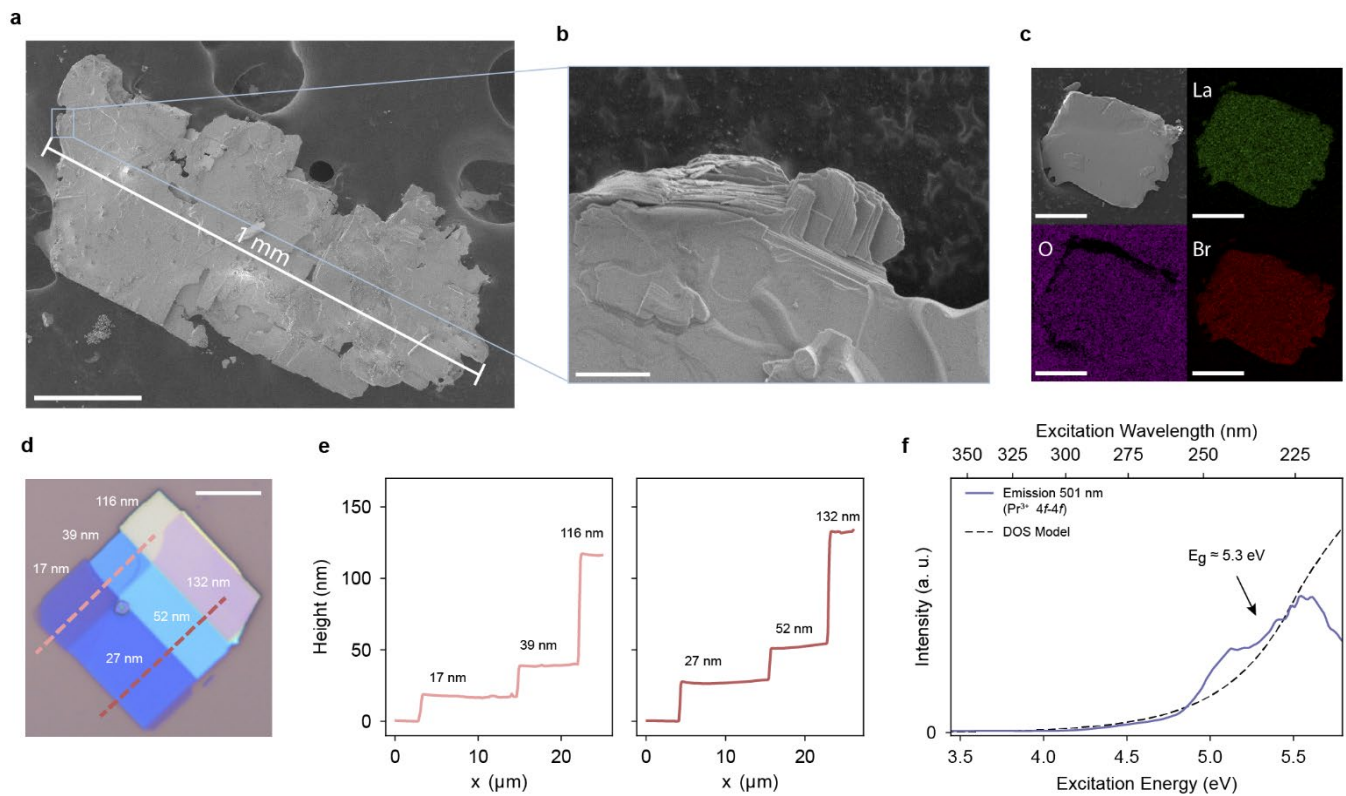

**Figure 2. Characterization of LaOBr crystals and exfoliated flakes. a,** SEM image of a LaOBr single crystal with a lateral dimension of 1mm obtained from the synthesis method used in this work. Scale bar: 200 µm. **b**, The magnified view of the crystal edge shows the terraced morphology indicative of a layered structure. Scale bar: 10 µm, **c**, EDS map on a single crystal of LaOBr, showing uniform distribution of elements across the whole crystal. Scale bar: 50 µm **d,** Flake of LaOBr on a SiO$_2$ substrate (270 nm), obtained following the standard tape exfoliation commonly employed for 2D materials and van der Waals heterostructure fabrication. Large flakes of tens of micrometers in width are easily obtained. The different colors of the flake terraces are related to different heights, with increasing values from dark blue to light blue (10 – 60 nm), towards yellow and pink (> 100 nm). Two shaded lines in light red and dark red represent two height measurements obtained by AFM. Scale bar: 10 µm. **e,** Height measurements of the flakes in (d) obtained by AFM, with colors corresponding to the light red and dark red shaded lines in (d). **f,** Photoluminescence excitation measurement on LaOBr crystals compared to a density of states (DOS) model[30], as further detailed in Supplementary Note 6. The emission wavelength is set to 501 nm (2.47 eV), corresponding to the 4f-4f transition of praseodymium impurity. We calculate the bandgap as 5.3 eV (Supplementary Note 6).

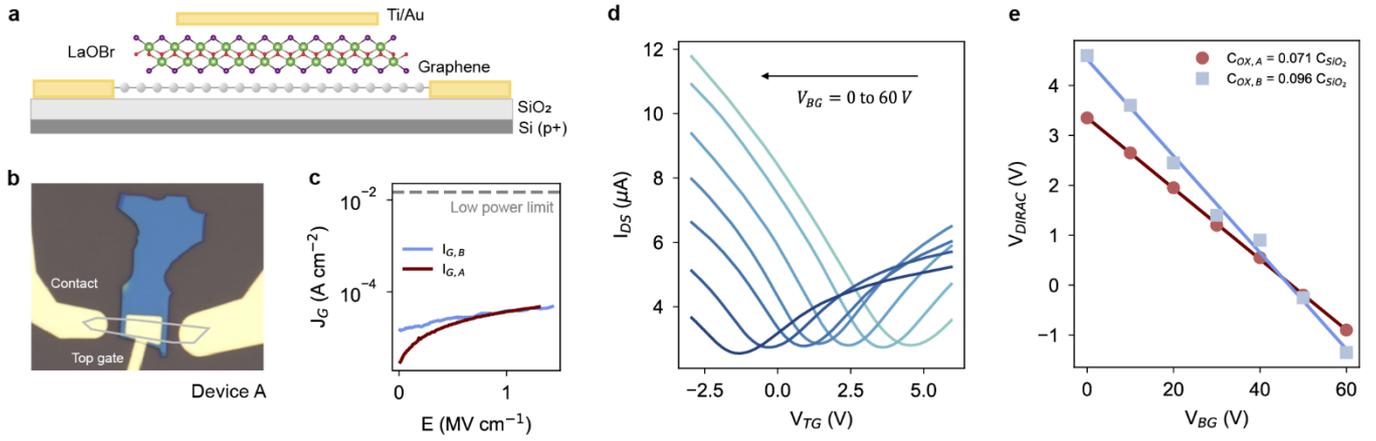

**Figure 3. LaOBr dielectric constant and leakage current. a,** Schematic of the graphene field-effect devices used for the estimation of the dielectric constant of LaOBr. Graphene is contacted by lateral Ti/Au electrodes, with a top LaOBr dielectric and a Ti/Au gate. The bottom dielectric is SiO$_2$ (270 nm) over highly p-doped Si. **b,** Optical micrograph of device A, showing the bulk LaOBr flake in blue and the graphene layer highlighted in grey. The Ti/Au lateral electrodes and top gate are labeled. **c,** Gate leakage current as a function of vertical electric field in the field-effect graphene devices A and B, showing consistent magnitudes of leakage ($< 10^{-4}$ A cm$^{-2}$) lying orders of magnitude below the low-power limit ($1.5 \cdot 10^{-2}$ A cm$^{-2}$). The back gate voltage is kept at $V_{BG} = 30\ V$ as a reference, while comparable results are obtained independently on $V_{BG}$. **d,** Drain-source current in device A as a function of the applied top gate voltage $V_{TG}$ for bottom gate voltages in the range $0 < V_{BG} < 60\ V$. The minima of the curves are obtained when the Fermi level lies at the graphene Dirac point position $V_{DIRAC}$. **e,** Change of $V_{DIRAC}$ with respect to $V_{BG}$ for devices A and B. The different slopes of the curves are related to the different thicknesses of the LaOBr layers, with 56 nm and 46 nm for devices A and B, respectively. The obtained capacitance ratios are highlighted in the legend of the plot. From these ratios, we use Eq. 1 to extract out-of-plane static dielectric constants of 8.6 and 9.4 for devices A and B, respectively, giving an average estimate of $\epsilon_{0,\perp} \simeq 9$.

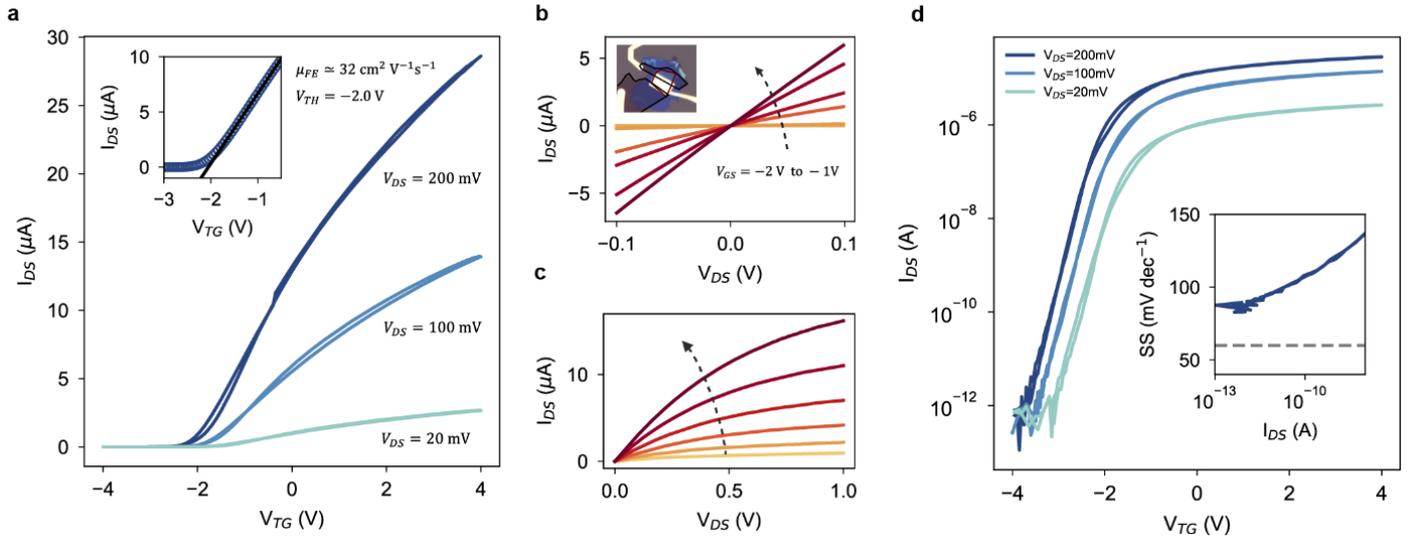

**Figure 4. MoS$_2$ field-effect transistor with LaOBr. a,** Drain-source current $I_{DS}$ of the 4-layer MoS$_2$ transistor with LaOBr as the gate dielectric as a function of the applied top gate voltage $V_{TG}$. Three different bias conditions $V_{DS}$ are reported and labeled on the curves. Inset: linear region of the transfer characteristic for $V_{DS} = 200$ mV, with an extracted threshold voltage $V_{TH} \simeq -2$ V and two-terminal field-effect mobility $\mu_{FE} \simeq 32$ cm$^2$ V$^{-1}$ s$^{-1}$. **b,** $I_{DS}$ as a function of $V_{DS}$ for gate voltages in the range $-2\,V < V_{GS} < -1\,V$. The range of bias is kept as $-0.1\,V < V_{DS} < 0.1\,V$ to highlight the linearity of the curves at small bias, an indication of ohmic contacts. Inset: optical micrograph of the transistor device, with the LaOBr flake (20 nm) in blue, while the few-layer graphite drain/source flakes and the MoS$_2$ flake are highlighted in black and red, respectively. The connections to the few-layer graphite flakes, as well as the top gate, are all made with Ti/Au. **c,** Output characteristics of the transistor for $-2\,V < V_{GS} < -1\,V$, showing a clear modulation from linear to saturation regions. **d,** Transfer characteristics from (a) plotted in logarithmic scale, highlighting the subthreshold and above-threshold regions for all biases. Inset: subthreshold swing from the local derivative $\partial V_{TG}/\partial I_{DS}$ as a function of $I_{DS}$, with minimum values around 85 mV dec$^{-1}$.

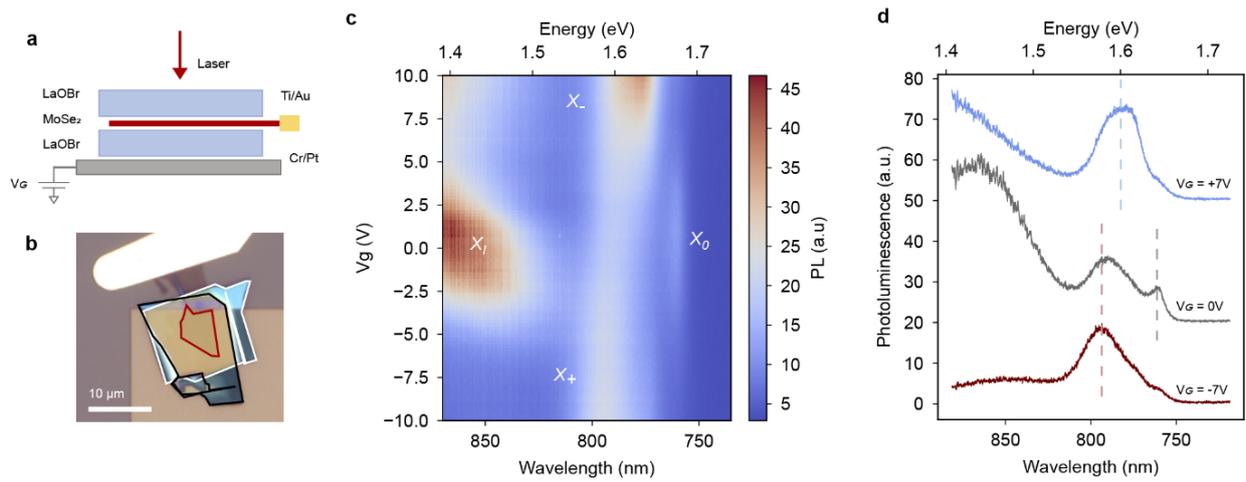

**Figure 5. Gate modulation of excitonic species in LaOBr-encapsulated monolayer MoSe2. a,** Schematic of the heterostructure based on monolayer MoSe$_2$ encapsulated with LaOBr on a Cr/Pt bottom gate. The semiconductor is contacted by a Ti/Au electrode. The structure is excited by a red laser (640 nm) focused at the diffraction limit with a power of 100 μW. **b,** Optical micrograph of the excitonic device, showing the monolayer MoSe$_2$ flake (red), together with the top (white) and bottom (black) LaOBr flakes. The Cr/Pt gate and the Ti/Au contact to MoSe$_2$ are clearly visible. Scale bar: 10 μm. **c,** Gate-dependent PL emission of the excitonic species in LaOBr-encapsulated monolayer MoSe$_2$. Neutral, charged (trions) and impurity-bound species are evidenced. **d,** Linecuts of PL emission at negative (red), zero (grey) and positive (blue) gate voltages. Neutral, negative and positive trion PL emission peaks are centered at approximately 1.63 eV, 1.60 eV, and 1.57 eV in energy. The broad impurity-bound exciton observed at lower energies is significantly quenched with increasing electrostatic doping, with predominant trion formation.